\documentstyle[aps,amsfonts,epsfig]{revtex}
\newcommand{\I}{\text{i}}
\newcommand{\E}{\text{e}}

\newcommand{\re}[1]{(\ref{#1})}
\newcommand{\sta}[1]{{}^\star\! #1}
\begin{document}
\draft
\twocolumn{
\title{Light Cone Condition for a Thermalized QED Vacuum}
\author{Holger Gies\thanks{E-mail address:
    holger.gies@uni-tuebingen.de}} 

\address{Institut f\"ur theoretische Physik\\
          Universit\"at T\"ubingen\\
      Auf der Morgenstelle 14, 72076 T\"ubingen, Germany}
\date{}
\maketitle
\begin{abstract}
  Within the QED effective action approach, we study the propagation
  of low-frequency light at finite temperature. Starting from a
  general effective Lagrangian for slowly varying fields whose
  structure is solely dictated by Lorentz covariance and gauge
  invariance, we derive the light cone condition for light propagating
  in a thermalized QED vacuum. As an application, we calculate the
  velocity shifts, i.e., refractive indices of the vacuum, induced by
  thermalized fermions to one loop. We investigate various temperature
  domains and also include a background magnetic field. While
  low-temperature effects to one loop are exponentially damped by the
  electron mass, there exists a maximum velocity shift of $-\delta
  v^2_{\text{max}}=\frac{\alpha}{3\pi}$ in the intermediate-temperature
  domain $T\sim  m$.
\end{abstract}
\section{Introduction}
As an application of the QED effective action approach at finite
temperature \cite{cox84}-$\!\!$\cite{gies98}, we investigate the
aberration of low-frequency photons from the usual vacuum light cone 
in the presence of a heat bath. The propagation of photons can
deviate from the classically free vacuum behavior when external
perturbations exert their influence on the vacuum polarization of the
photon. In a perturbative language, virtual processes, e.g., the
electron-positron loop, confer the properties of the participant
particles to the photon. By this means, the photon can therefore
acquire, for instance, a ``size'' (of the order of the Compton
wavelength of the participant particle) and a ``charge distribution''.
Finally, external perturbations, e.g., electromagnetic or
gravitational fields, cavity-like boundary conditions, or a heat bath,
can interact with these induced photon properties. The resulting
modified photon propagation can be expressed in terms of a light cone
condition which describes the deformation of the usual vacuum light
cone as observed by the photons with frequency $\omega$ lower than the
electron mass $m$. A variety of these effects have been under
consideration in recent years (see, e.g., \cite{ditt98} and references
cited therein). 

Concentrating on the deviation of the speed of light from the
unmodified-vacuum velocity $c$ (=1 in our units) by thermal
effects, the velocity shift\footnote{The velocity shift equals minus
  the shift of the refractive index associated with the modified
  vacuum.} in the low-temperature region, $T\ll m$, where $m$ denotes
the electron mass, yields $\delta v^2\sim -T^4/m^4$.  This important
result can be obtained from the two-loop polarization tensor, in which
the radiative photon within the fermion loop is considered to be
thermalized \cite{tarr83}, \cite{lato95}. The same result has been
found by an effective one-loop calculation employing the
Heisenberg-Euler Lagrangian \cite{bart90}\footnote{In this reference,
  the correct numerical result was found for the first time;
  furthermore, the similarities between the present problem and the
  Scharnhorst effect \cite{scha90}, i.e., light propagation in a
  Casimir vacuum, have been pointed out.}. The latter approach was
also put forward in \cite{ditt98}. Therein, the equations of motion of
the effective Lagrangian have been used to derive the light cone
condition, and the effects of finite temperature have been introduced
by evaluating thermal vacuum expectation values of the field
quantities. The high-temperature domain has also been studied in
\cite{ditt98} by taking the effects of additionally thermalized
fermions approximately into account.

Since the effective action provides us with a convenient interface
between the full quantum theory and classical field theory, it is the
appropriate tool with which to describe electromagnetic waves of
low frequency ($\omega\ll m$), which are the low-energy degrees of
freedom of QED. The modified, e.g., thermalized, vacuum is viewed as a
medium with optical properties which exert an influence on the
propagation of a plane wave. In the present work, we rigorously extend
the effective action approach to the light cone condition as applied in
\cite{ditt98} to the finite-temperature case. For this, the dependence
of the effective action on the complete set of gauge and
Lorentz invariants of the given situation has to be considered;
contrary to the zero-temperature case involving the two 
standard invariants ${\cal F}$ and ${\cal G}$, we have to take two
further invariants into account which arise from an additional
four-vector associated with the heat bath. This is compulsory for an
exact treatment of the problem in the effective action approach, and
has not been considered in earlier works. The only input for deducing
the light cone condition for low-frequency waves will be a general
Lagrangian whose dependence on electromagnetic field and temperature
is solely dictated by Lorentz covariance and gauge
invariance. Therefore, the derivation of the light cone condition does
not rely on perturbation theory or a perturbative expansion of the
Lagrangian.  

As an application for our formalism, we fall back on perturbation
theory and employ the thermal one-loop effective action of QED which
allows for an investigation of light propagation in a low- and an
intermediate-temperature domain, $T\sim m$. It is a priori clear that
the low-temperature velocity shift $\sim T^4$ will not be found to
this order of calculation, since it is a two-loop effect in an exact
field theoretic framework\footnote{In \cite{ditt98}, the loop counting
  refers to the order of the Heisenberg-Euler Lagrangian employed
  during the approximation process for thermalizing the fermions and
  therefore differs from the present one.}. Unfortunately, a full
two-loop calculation of the thermal QED effective action has not yet
been performed. At intermediate temperature, $T\sim m$, the one-loop
contribution is expected to be the dominant one. Apart from soft
logarithmic corrections, we find a maximum velocity shift which
reduces the light velocity by $\delta
v_{\text{max}}^2=-\frac{\alpha}{3\pi}$.

Our results do not apply to values of temperature at which the physics
of the $e^+e^-$-gas is dominated by plasma effects, $T\gg m$. Then,
signal propagation will completely be determined by plasma modes which
are not considered in our approach. Especially, waves of
low-frequency cease to propagate, since the appearance of a plasma
mass serves as a cutoff for low frequencies.

\section{Effective Action at Finite Temperature}
The low-energy effective action of QED for slowly varying fields at
finite temperature and zero density can only depend on a small set of
invariants. In particular, a dependence of the effective action on the
(infinitely many) invariants involving derivatives of the fields can
be neglected if the typical scale associated with the variation of the
fields in spacetime is much larger than the Compton wavelength of the
electron ($\lambda_{\text{c}}=1/m$). Concerning light propagation,
this is a reasonable assumption as long as the frequency of the
propagating wave is much smaller than the electron mass ($\omega\ll
m$) and possible additional background fields are almost homogeneous.  

In the sense of relativistic equilibrium thermodynamics, the Lorentz
covariant and gauge invariant building blocks of these invariants are
given by the field strength tensor and its dual, $F^{\mu\nu},
\sta{F}^{\mu\nu}$, and by the heat-bath vector $n^\mu$.  The latter is
on the one hand characterized by the value of its invariant scalar
product\footnote{We employ the metric {\em g}=diag(-1,1,1,1).}, $n^\mu
n_\mu=-T^2$, where $T$ denotes the heat-bath temperature, and on the
other hand related to the heat-bath 4-velocity $u^\mu$ via the
invariant parameter $T$: $n^\mu= T\, u^\mu$.

From these objects, we can construct the following set of invariants,
which represents a conventional choice:
\begin{eqnarray}
{\cal F}&=&\frac{1}{4}F^{\mu\nu}F_{\mu\nu}=\frac{1}{2}
\bigl({\mathbf{B}^2-\mathbf{E}^2}\bigr), \nonumber\\
{\cal G}&=&\frac{1}{4}\sta{F}^{\mu\nu}F_{\mu\nu}
=-{\mathbf{E\cdot B}}, \nonumber\\
{\cal E}&:=&\frac{1}{T^2}\bigl(n_\alpha F^{\alpha\mu}\bigr)
  \bigl(n_\beta  F^{\beta}{}_\mu\bigr)
\equiv\bigl(u_\alpha F^{\alpha\mu}\bigr)  \bigl(u_\beta
  F^{\beta}{}_\mu\bigr), \nonumber\\ 
T&=&\sqrt{-n^\mu n_\mu}.\label{1.1}
\end{eqnarray}
Incidentally, parity invariance requires the action to be an even
function of ${\cal G}$ and prohibits the generation of a
``Chern-Simons''-like term $\sim u_\mu \sta{F}^{\mu\nu} A_\nu$.
Without loss of generality, we assume ${\cal G}>0$ for reasons of
uniqueness. The convenience of the choice of ${\cal E}$ becomes clear
in the heat-bath rest frame, where we find ${\cal
  E}={\mathbf{E}}^2$. It is easy to check that the maximum number of
classical invariants obeying these symmetry requirements is four (see,
e.g. Sec. III of Ref. \cite{gies98}); the
set of Eq. \re{1.1} is thus complete.

From these symmetry considerations, we conclude that the low-energy
effective Lagrangian of QED for slowly varying fields at finite
temperature can generally be written as\footnote{Apart from the
  implicit (and soft) spacetime dependence of the fields, the system
  under consideration is homogeneous in space and time (thermal
  equilibrium). Therefore, the effective Lagrangian cannot depend
  explicitly on the coordinates $x^\mu$. Although the
  finite-temperature case closely resembles the case of a Casimir
  vacuum, the latter is not invariant under translations orthogonal to
  the Casimir plates; hence, the effective Lagrangian for the Casimir
  vacuum can and, in fact, does depend explicitly on the spacetime
  coordinates \cite{scha90}. In order to perform the considerations of
  the present work for the Casimir vacuum, we thus had to include
  another invariant of the form $x^\mu n_\nu$ where $n^\mu$ denotes
  the normal vector associated with the Casimir plates.}:
\begin{equation}
{\cal L}={\cal L}\bigl({\cal F},{\cal G},{\cal
  E},T\bigr). \label{1.1a}
\end{equation}
In other words, we are dealing with the local part (no derivatives of
the fields) of the complete non-local effective action (involving
infinitely many derivative terms of the fields). Of course, even the
local part of the exact effective action incorporating the
contributions of all loops is presently out of reach. Hence in
practical applications, we will rely on a perturbative expansion of
Eq. \re{1.1a}; in fact, only the lowest order has been identified up
to now. The derivation of this one-loop QED effective action for
arbitrary constant electromagnetic fields at finite temperature has
required more effort than the purely magnetic case \cite{ditt79+}. The
first comprehensive study has been elaborated by Elmfors and
Skagerstam \cite{elmf94} employing the real-time formalism. For our
purposes, we make use of the representation of the effective action as
given in \cite{gies98}. Therein, the field and temperature dependence
is expressed in terms of the complete set of invariants as given in
Eq. \re{1.1}.

Furthermore, introducing the abbreviations
\begin{eqnarray}
a&:=&\left( \sqrt{{\cal F}^2+{\cal G}^2}+{\cal
    F}\right)^{\frac{1}{2}}, \nonumber\\ 
b&:=&\left( \sqrt{{\cal F}^2+{\cal G}^2}-{\cal
    F}\right)^{\frac{1}{2}}, \label{1.2}
\end{eqnarray}
the thermal contribution to the one-loop effective Lagrangian at zero
density reads \cite{gies98}
\begin{eqnarray}
{\cal L}^{1T}&=&\frac{1}{4\pi^2} \int\limits_0^\infty \frac{ds}{s^3}
  \,\E^{-\I m^2 s}\, eas\cot (eas) \,ebs\coth (ebs) \label{1.3}\\
&&\qquad\qquad\times\sum_{n=1}^\infty (-1)^n \E^{\I
  h(s)\frac{n^2}{4T^2}}, \nonumber
\end{eqnarray}
whereby the exponent is given by
\begin{equation}
h(s)=\frac{b^2-{\cal E}}{a^2+b^2}\, ea \cot eas 
     +\frac{a^2+{\cal E}}{a^2+b^2}\, eb\coth ebs. \label{1.3a}
\end{equation}
The complete effective Lagrangian then consists of ${\cal L}={\cal
  L}_{\text{cl}} +{\cal L}^1+{\cal L}^{1T} +{\cal L}_\gamma$, where
${\cal L}_{\text{cl}}=-{\cal F}$ denotes the classical Lagrangian and
${\cal L}^1$ represents the well-known one-loop part \cite{schw51} at
zero temperature, i.e., the Heisenberg-Euler-Schwinger Lagrangian. For
later use, we also incorporated the free photonic contribution ${\cal
  L}_\gamma$ which is equal to minus the free energy density of a free
photon gas (Stefan-Boltzmann law), ${\cal
  L}_\gamma(T)=\frac{\pi^2}{45} T^4$. Incidentally, only the
zero-temperature one-loop part ${\cal L}^1$ is affected by
renormalization; for fixing the counter-terms, we require the complete
Lagrangian to meet with the pure classical part in the weak-field
limit (at $T=0$).

\section{Light Cone Condition}

The system under consideration is a propagating plane wave field in
the presence of a vacuum that is modified by an external constant
electromagnetic field at finite temperature and zero density.
Following \cite{ditt98}, we assume the propagating wave to be of low
frequency and neglect any vacuum modifications caused by the
propagating light itself. In other words, we neglect dispersive
effects and exclude non-linear self-interactions of the wave. Thus, we
can linearize the field equations with respect to this plane wave field.

Within the framework of these assumptions, the field equations are
obtained from the Euler-Lagrange equations of motion of the effective
thermal QED Lagrangian:
\begin{eqnarray}
0&=&\!2\partial_\mu \frac{\partial {\cal L}}{\partial F_{\mu\nu}}
%\nonumber\\
=\!2\partial_\mu\!\!\left[\partial_{\cal F}{\cal L} \frac{\partial {\cal
    F}}{\partial F_{\mu\nu}}\!+\partial_{\cal G}{\cal L} \frac{\partial
  {\cal G}}{\partial F_{\mu\nu}}\! +\partial_{\cal E}{\cal L} \frac{\partial
  {\cal E}}{\partial F_{\mu\nu}}\right]\!. \nonumber\\
&&\label{2.1} 
\end{eqnarray}
The Lagrangian ${\cal L}$ that we are going to insert into Eq. \re{2.1}
contains the classical part as well as the zero-temperature and
thermal one-loop contributions: ${\cal L}={\cal L}_{\text{cl}} +{\cal
  L}^1+{\cal L}^{1T}$; however, the following derivation of the light
cone condition does not rely on any perturbative approximation of
${\cal L}$. It is only necessary that the field dependence of the
Lagrangian be completely contained in a set of three linearly
independent invariants for which we take the standard invariants
${\cal F}$, ${\cal G}$ and the invariant ${\cal E}$ as defined in
Eq. \re{1.1}. Note that the differentiation with respect to
$F_{\mu\nu}$ has to be performed with regard to its anti-symmetry
properties.  

After moving the space-time derivative to the right employing the
Bianchi-identity $\partial_\mu \sta{F}^{\mu\nu}=0$, we may write
Eq. \re{2.1} in the form
\begin{equation}
0=\partial_{\cal F} {\cal L}\, \partial_\mu F^{\mu\nu}+\partial_\mu
  F^{\alpha\beta}\, \left(\frac{1}{2} M^{\mu\nu}_{\alpha\beta}
  \right), \label{2.2}
\end{equation}
where $M^{\mu\nu}_{\alpha\beta}$ denotes the tensor,
\begin{eqnarray}
&&\frac{1}{2} M^{\mu\nu}_{\alpha\beta} 
  =4\partial_{\cal E} {\cal L}\,u^{[\mu} u_{[\alpha} 
                                  \delta_{\beta]}^{\nu]}
\nonumber\\
&&\quad\! 
+F^{\mu\nu} 
  \left[\frac{\partial_{\cal F}^2{\cal L}}{2} F_{\alpha\beta} 
    +\frac{\partial_{{\cal F}{\cal G}}{\cal L}}{2}
      \sta{F}_{\alpha\beta}  
    +2\partial_{{\cal F}{\cal E}}{\cal L}\, 
    u_{[\alpha} (u\mathsf{F})_{\beta]} \right] \nonumber\\
&&\quad\!
+\sta{F}^{\mu\nu} 
  \left[\frac{\partial_{{\cal F}{\cal G}}{\cal L}}{2}  
         F_{\alpha\beta} 
    +\frac{\partial_{{\cal G}}^2{\cal L}}{2}\sta{F}_{\alpha\beta} 
    +2\partial_{{\cal G}{\cal E}}{\cal L}\, 
      u_{[\alpha} (u\mathsf{F})_{\beta]} \right] \nonumber\\
&&\quad \!
+4u^{[\mu} (u{\mathsf{F}})^{\nu]} \!\!
  \left[\frac{\partial_{{\cal F}{\cal E}}{\cal L}}{2}  
          F_{\alpha\beta} 
    +\!\frac{\partial_{{\cal G}{\cal E}}{\cal L}}{2} 
          \sta{F}_{\alpha\beta} 
    +\!2\partial_{{\cal E}}^2{\cal L}\, 
      u_{[\alpha} (u{\mathsf{F}})_{\beta]}\! \right]\!\!, \nonumber\\
&&\label{2.2a}
\end{eqnarray}
where we employed matrix notation, $({\mathsf{F}}k)^\nu\equiv
F^\nu{}_\alpha k^\alpha$, or $(ku)=k^\alpha u_\alpha$, and a
prescription for anti-symmetrizing tensor indices:
$A^{[\mu\nu]}=\frac{1}{2} \bigl(A^{\mu\nu} -A^{\nu\mu}\bigl)$. Note
that the tensor $M^{\mu\nu}_{\alpha\beta}$ is anti-symmetric with
respect to its upper as well as its lower indices, and contains first
and second derivatives of the Lagrangian with respect to the three
invariants. As mentioned above, there is no explicit dependence of
${\cal L}$ on the coordinates because of homogeneity in spacetime.
Equation \re{2.2} represents the effective field equations for slowly
varying fields in a thermalized QED vacuum. It replaces the linear
Maxwell equations in vacuum $\partial_\mu F^{\mu\nu}=0$, defining a
new theory of ``classical'' non-linear electrodynamics.

Following the techniques of \cite{ditt98}, we adapt these equations to
the case of a propagating plane wave: first, we decompose the
electromagnetic field strength into the field strength of the plane
wave field $f^{\mu\nu}$ and an additional homogeneous background field
$F^{\mu\nu}$: $F^{\mu\nu} \to F^{\mu\nu}+ f^{\mu\nu}$. Now derivatives
of the field act only on the plane wave part: $\partial_\mu
(F^{\alpha\beta} +f^{\alpha\beta})=\partial_\mu f^{\alpha\beta}$.
Transition to Fourier space leads us to the replacement $-\I
\partial_\mu f^{\alpha\beta} \to k_\mu f^{\alpha\beta}$, where $k_\mu$
denotes the wave vector of the plane wave field $f^{\alpha\beta}$.
The latter is proportional to $f^{\alpha\beta} \sim k^\alpha
\epsilon^\beta -k^\beta \epsilon^\alpha$, whereby $\epsilon^\alpha$
represents a polarization vector of the plane wave. Furthermore, we
adopt the Lorentz gauge condition for the plane wave field and sum
over polarization states; the latter step is, on the one hand,
convenient, and, on the other hand, reasonable for a purely
thermalized vacuum which is naturally isotropic. In the case of an
additional background field, we lose information about effects of
birefringence by summing over polarization states
(cf. later). Finally, we arrive at the desired light cone condition:
\begin{equation}
0=2\, \partial_{\cal F}{\cal L}\, k^2+2 \left( \frac{1}{2}
  M^{\mu\nu}_{\alpha\nu} \right) \, k_\mu k^\alpha. \label{2.3}
\end{equation}
Since we have linearized with respect to the propagating plane wave
field, the field quantities appearing in Eq. \re{2.3} now describe an
externally applied background field only. Following summation over
polarization states, the field strength tensor of the plane wave field
has completely dropped out of the formula. This corresponds to our
approximation of neglecting vacuum modifications caused by the plane
wave field itself. In this sense, the plane wave field resembles a
small test charge of classical electrodynamics.

Evaluating the contractions of $M^{\mu\nu}_{\alpha\nu}$ with $k_\mu
k^\alpha$, we find the light cone condition of a soft plane wave
propagating in an external field at finite temperature in its final
form,
\begin{eqnarray}
0\!&=&\!\bigl(\partial_{\cal F}{\cal L}\!+{\cal G}\partial_{{\cal F}
  {\cal G}}{\cal L}\!-\!{\cal F}\partial^2_{\cal G}{\cal L} \bigr)k^2\!
  +\case{1}{2} \bigl( \partial^2_{\cal F} \!+\partial^2_{\cal G}\bigr)
  {\cal L} ({\mathsf{F}}k)^\nu({\mathsf{F}}k)_\nu \nonumber\\
&&-\partial_{{\cal E}}{\cal L}\, k^2+2\bigl[ \partial_{\cal E}{\cal
  L}+{\cal E}\partial^2_{\cal E}{\cal L}\bigr]\, 
  (ku)^2\label{2.4}\\
&&+2\partial_{{\cal F}{\cal E}}{\cal L}\, (ku)({\mathsf{F}}k)^\nu
  ({\mathsf{F}}u)_\nu +2\partial_{{\cal G}{\cal E}}{\cal L}\,
  (ku)({\sta{\mathsf{F}}}k)^\nu  ({\mathsf{F}}u)_\nu \nonumber\\
&& +2(\partial_{{\cal F}{\cal E}} {\cal L} -\partial_{{\cal E}}^2{\cal
  L})\, (u{\mathsf{F}}k)^2
  +2\partial_{{\cal G}{\cal E}} {\cal L}\, (u{\mathsf{F}}k)
  (u{\sta{\mathsf{F}}}k). 
  \nonumber
\end{eqnarray}
Let us stress once more that the field invariants ${\cal F},{\cal
  G},{\cal E}$ in this equation characterize an externally applied
background field only. The first line in Eq. \re{2.4} corresponds to
the purely field-modified light cone condition as derived in
\cite{ditt98}, which can conveniently be expressed in terms of the
energy-momentum tensor (or its vacuum expectation value) of the
electromagnetic field.

In the following, we simply confine ourselves to the case of vanishing
external field: $F^{\mu\nu},{\cal F},{\cal G},{\cal E}\to 0$. In this
limit, the light cone condition becomes extremely simplified, and,
except for the term linear in ${\cal F}$ which is filtered out by
$\partial_{\cal F}{\cal L}|_{{\cal F},{\cal G},{\cal E}=0}$, only
the first term in square brackets in Eq. \re{2.4} survives:
\begin{equation}
0=(\partial_{\cal F}{\cal L}-\partial_{\cal E}{\cal L})\, k^2
  +2\partial_{\cal E}{\cal L}\, (ku)^2, 
  \quad
  \text{for}\quad {\cal F},{\cal G},{\cal E}\to 0. \label{2.5}
\end{equation}
Introducing the phase velocity $v=\frac{k^0}{|{\mathbf{k}}|}$, which
is identical to the group velocity in the low-frequency limit, we may
rewrite the light cone condition in the heat-bath rest frame
($(ku)^2=(k^0)^2$) in terms of the squared velocity,
\begin{equation}
v^2=\frac{1}{1+\frac{2\, \partial_{\cal E}{\cal L}}{(-\partial_{\cal F}
    {\cal L} +\partial_{{\cal E}}{\cal L})}}. \label{2.6}
\end{equation}
In the present case of light propagation in a thermalized vacuum, it
is physically reasonable to expect that $v\leq 1$ holds rigorously,
since the propagating photons do not only virtually interact with the
usual vacuum modes (zero-point fluctuations) but also with the
thermally excited modes. This may be interpreted as a kind of
``resistance'' implying a decrease of the propagation velocity.   
In order to maintain $v\leq 1$, the fraction in the denominator of
Eq. \re{2.6} should always be positive. Note that this statement is
independent of any loop approximation.

Now we turn to perturbation theory, where ${\cal L}$ can be assumed to
consist of the classical term ${\cal L}_{\text{cl}}=-{\cal F}$ plus
quantum correction terms such as those discussed in Sec. II. The
latter are considered to be small compared to the classical one;
hence, Eq. \re{2.6} to lowest order in the perturbation simplifies to:
\begin{equation}
v^2\simeq 1-2\, \partial_{\cal E}{\cal L}. \label{2.6aa}
\end{equation}
As will be demonstrated below, this formula describes sufficiently the
thermally induced velocity shift to one loop for reasonable values of
temperature. 

For the remainder of the section, we will transcribe the light cone
condition Eq. \re{2.5} into a form which has proved useful for gaining
intuitive insight into the problem of light propagation in modified
vacua \cite{ditt98}. For this, we introduce the vacuum expectation
value of the energy-momentum tensor,
\begin{equation}  
\langle T^{\mu\nu}\rangle ={\cal L}\, g^{\mu\nu} + T\, \partial_T
  {\cal L}\, u^\mu u^\nu, \label{2.6a}
\end{equation}
which can be obtained by varying the effective action with respect to
the metric $g_{\mu\nu}$, and treating the variable $T$ as
$T=\sqrt{-n^\mu g_{\mu\nu} n^\nu}$. Equation \re{2.6a} is the {\em
  finite-temperature} analogue of Eq. (19) of \cite{ditt98} where the
vacuum expectation value of $T^{\mu\nu}$ for an {\em electromagnetic}
background has been considered. Solving Eq. \re{2.6a} for $u^\mu
u^\nu$, we may substitute this into Eq. \re{2.5} and arrive at:
\begin{equation}
k^2=Q_T\, \langle T^{\mu\nu} \rangle\, k_\mu k_\nu, \label{2.6b}
\end{equation}
where the so-called {\em effective action charge} \cite{ditt98} $Q_T$
for the present system is given by\footnote{In \cite{ditt98}, the name
  ``effective action charge'' derives from the resemblance between the
  corresponding equation to Eq. \re{2.6c} for electromagnetic fields
  and the Poisson equation, $2Q:=(\partial^2_{{\cal
      F}}+\partial^2_{{\cal G}}){\cal L}$. This accidental feature
  does not hold in the present case; however, we stick to this
  nomenclature, since essential properties as well as the underlying
  physical picture also hold here.}:
\begin{equation}
Q_T=
\frac{\frac{2}{T}\frac{\partial_{\cal E} {\cal L}}{\partial_T {\cal
      L}} } 
   {\left(-\partial_{{\cal F}}{\cal L}+ 
          \partial_{\cal E} {\cal L}+\frac{2}{T} \frac{\partial_{\cal 
         E} {\cal L}}{\partial_T {\cal L}}\,  {\cal L}\right) }. 
\label{2.6c} 
\end{equation}
From Eq. \re{2.6b}, we can read off that the deformation of the light
cone is determined by the (renormalized) energy density of the
modified vacuum. The proportionality factor $Q_T$ depends on the
parameters of the system via the effective Lagrangian. Under the
reasonable assumption that the deformation of the light cone is
bounded even in the high-temperature domain, one can conclude that the
factor $Q_T$ has to decrease sufficiently fast for increasing 
temperature. Therefore, we expect $Q_T$ to be locally
(charge-like) distributed in the parameter space of temperature. This
justifies the nomenclature {\em effective action charge}. In fact, the
one-loop results will confirm this picture.  

\section{One-Loop Results}
As an application, we insert the one-loop effective action into the
light cone condition. Let us first consider a thermalized vacuum with
vanishing background fields. Therefore, the velocity shift stems
from the thermal contribution ${\cal L}^{1T}$ only, and we need
$\partial_{\cal E}{\cal  L}=\partial_{\cal E}{\cal L}^{1T}$ in the
zero-field limit: 
\begin{equation}
\partial_{\cal E}{\cal L}^{1T}\stackrel{\re{1.3}}{=}-\frac{\alpha}{3\pi}
\sum_{n=1}^\infty (-1)^n\, \left(\frac{m}{T} n\right)\, K_1 \left(
  {\scriptstyle \frac{m}{T} n} \right). \label{2.7}
\end{equation}
Regarding Eq. \re{2.6a}, we also need $\partial_{\cal F}{\cal L}$; for
this, note that the zero-temperature one-loop part is renormalized in
such a way that the term linear in ${\cal F}$ vanishes in order to
recover the pure Maxwell theory in the weak-field limit. Hence, it is
only the thermal contribution ${\cal L}^{1T}$ which provides for an
additional term linear in ${\cal F}$: $\partial_{\cal F}{\cal L}=-1+
\partial_{\cal F}{\cal L}^{1T}$, where the $(-1)$ stems from the
Maxwell part ${\cal L}_{\text{cl}}$. In the zero-field limit, we get:
\begin{equation}
\partial_{\cal F}{\cal L}^{1T}=-\frac{4\alpha}{3\pi}\sum_{n=1}^\infty
(-1)^n K_0\bigl({\scriptstyle \frac{m}{T} n} \bigr). \label{2.7a}
\end{equation} 
For Eqs. \re{2.7} and \re{2.7a}, we employed the fact that summation
and proper-time integration in Eq. \re{1.3} are interchangeable in the
zero-field limit.  The proper-time integrals can then be identified as
a representation of the modified Bessel function $K_1$ and $K_0$
\cite{GR}. 

In the limiting cases of low, intermediate and high temperature, Eqs.
\re{2.7} and \re{2.7a} can be expanded in terms of the parameter
$\frac{T}{m}$. Let us first consider Eq. \re{2.7a}: at low
temperature, only the first term of the sum must be taken into
account, which leads to
\begin{equation}
\partial_{\cal F}{\cal L}^{1T}(T\ll m)\simeq \frac{2\alpha}{3\pi}
\,\sqrt{\frac{2\pi T}{m}}\, \E^{-\frac{m}{T}}. \label{2.8a}
\end{equation}
For an intermediate- or high-temperature expansion, the infinite sum
must be completely taken into account; this can be achieved employing
the techniques proposed in the appendix B of \cite{ditt98}. The result
for Eq. \re{2.7a} at $T\sim m$ is
\begin{equation}
\partial_{\cal F}{\cal L}^{1T}=\frac{2\alpha}{3\pi} \left
  [ (0.666\dots) +(0.814\dots) \ln\frac{T}{m} \right], \label{2.8b}
\end{equation}
where the numbers stem from pure integrals over analytic
functions. Incidentally, the expansion for $T\gg m$ is formally
identical to Eq. \re{2.8b} with the factor of 0.814\dots replaced by
1 as found in \cite{elmf98}. 

For our purpose of investigating low- and intermediate temperature
domains, we observe that $\partial_{\cal F}{\cal L}^{1T}(T)={\cal
  O}(\alpha/\pi)\ll 1$. For the calculation of the velocity
shifts to order $\alpha/\pi$, it is thus sufficient to employ
Eq. \re{2.6aa}, i.e., neglecting $\partial_{\cal F}{\cal L}^{1T}$
compared to $\partial_{\cal F}{\cal L}_{\text{cl}}=-1$.  

Turning to the low-temperature expansion of Eq. \re{2.7}, we have to
take into account only the first term of the sum:
\begin{equation}
\partial_{\cal E}{\cal L}^{1T}(T\ll m)\simeq \frac{\alpha}{3\pi}
\,\sqrt{\frac{\pi}{2}} \sqrt{\frac{m}{T}}\,
\E^{-\frac{m}{T}}. \label{2.8}
\end{equation}
Inserting this into Eq. \re{2.6aa}, we find for the squared velocity at
low temperature an exponentially decreasing modification:
\begin{equation}
v^2(T\ll m)\simeq 1-\frac{\alpha}{3}\, \sqrt{\frac{2}{\pi}}\,
\sqrt{\frac{m}{T}}\, \E^{-\frac{m}{T}}. \label{2.9}
\end{equation}
The high-temperature expansion of Eq. \re{2.7} can be worked out along
the lines of App. B of \cite{ditt98}, leading to
\begin{equation}
\partial_{\cal E}{\cal L}^{1T}(T\gg m)\simeq  \frac{\alpha}{6\pi} \, \left(
  1-k_2 \frac{m^2}{T^2} +{\cal O}\left( {\scriptstyle \frac{m^4}{T^4}}
  \right) \right), \label{2.10}
\end{equation}
where $k_2$ denotes the number $k_2=0.213139\dots$ and arises from a
parameter-independent integral over analytic functions during the
expansion. The light cone condition then yields for the squared
velocities at high temperature:
\begin{equation}
v^2(T\gg m)\simeq 1-\frac{\alpha}{3\pi} \left( 1+k_2 \frac{m^2}{T^2}
\right)+\dots, \label{2.11}
\end{equation}
which implies a maximum velocity shift of $\delta
v_{\text{max}}^2= -\frac{\alpha}{3\pi}\simeq -\frac{1}{1291}$. 

It should, however, be noted that an expansion for $T/m\gg 1$ is a
formal trick to extract analytical results. For identifying the
values of temperature to which the light cone condition Eq. \re{2.4}
is applicable, 
we notice that the frequency of the plane wave should, on the one
hand, be smaller than the electron mass in order to justify the
assumption of slowly varying fields, and on the other hand, be larger
than the plasma frequency in order to ensure the existence of such a
propagating mode:
\begin{equation}
\omega_{\text{p}} \ll \omega \ll m. \label{2.12}
\end{equation}
For the plasma frequency corresponding to the Debye screening mass, we
employ the representation found in \cite{gies98}:
\begin{equation}
\omega^2_{\text{p}}=-8 \frac{\alpha}{\pi}\, m^2 \sum_{n=1}^\infty
(-1)^n\, K_2({\scriptstyle \frac{m}{T}} n). \label{2.13}
\end{equation}
The maximum value of temperature up to which the low-frequency assumption
can formally be justified is determined by
$\omega_{\text{p}}(T_{\text{max}})=m$. Numerically, one finds
$T_{\text{max}}/m \simeq 5.74\dots$. Of course, this is just a formal
value; in order to obtain reasonable results, i.e., in order to
satisfy relation \re{2.12}, the actual temperature
should be kept smaller than this maximum value. 

Nevertheless, the numerical results given below confirm that the
formal expansion for $T/m\gg 1$ is already appropriate for $T\sim m$,
which justifies extracting physical conclusions from this analytical
procedure. 

For the remainder of the section, we additionally allow for a further
perturbation of the vacuum in form of an external weak magnetic field,
$B\ll B_{\text{cr}}=\frac{m^2}{e}$. In terms of the invariants, 
purely magnetic fields imply that ${\cal G},{\cal E}\to 0$, ${\cal
  F}\to \frac{1}{2} B^2$, where the latter expression is understood to
be valid in the heat-bath rest frame. Due to the simple form of the
four-velocity vector $u^\mu=(1,0,0,0)$ in this special frame, terms
containing $F^{\mu\nu}u_\nu$ vanish, since this product is
proportional to the electric field, which is assumed to be zero. 

Taking these considerations into account, the light cone condition
\re{2.4} is reduced to:
\begin{eqnarray}
0\!&=&\!\bigl(\partial_{\cal F}{\cal L}\!-\partial_{{\cal E}}{\cal
  L}\!-{\cal F}\partial^2_{\cal G}{\cal L} \bigr)k^2\! 
+\case{1}{2} \bigl( \partial^2_{\cal F} \!+\partial^2_{\cal G}\bigr)
  {\cal L} ({\mathsf{F}}k)^\nu({\mathsf{F}}k)_\nu \nonumber\\
&&+ 2\partial_{\cal E}{\cal L}\, 
  (ku)^2.\label{2.14}
\end{eqnarray}
We will restrict the following investigations to experimentally more
accessible parameter values, i.e., weak fields and low temperature
(compared to the electron mass). Here we would like to answer the
question of whether there is an additionally induced velocity shift
for the case of combined magnetic {\em and} thermal vacuum
modifications. For each single case, the velocity shifts are known;
e.g., the polarization-summed velocity shift of light propagating in a
magnetic background field is given by \cite{adle71}:
\begin{equation}
\delta v_B^2 =-\frac{22}{45} \frac{\alpha^2}{m^4}\, B^2\, \sin^2
\Theta, \label{2.14a}
\end{equation}
where $\Theta$ denotes the angle between the propagation direction and
the magnetic field. The velocity shift for the purely thermal part is
found in Eq. \re{2.9}.

In order to find a contribution to the velocity shift for the combined
case apart from the trivial sum of the single cases, we need an
appropriate expansion of the thermal one-loop Lagrangian ${\cal
  L}^{1T}$ for weak fields and low temperature. This expansion has
been obtained in a systematic way in \cite{elmf98}; the same result
can be found by a direct Taylor expansion of Eq. \re{1.3}. In the
desired limit, the required terms which are quadratic in the field
invariants and dominant at low temperatures read: 
\begin{eqnarray}
&&\Delta{\cal L}_{BT}({\cal F},{\cal G},{\cal E},T)\label{2.14b}\\
&&\quad=\frac{\alpha^2}{45} \left[ \frac{8{\cal F}{\cal E}-8{\cal
      F}^2-13{\cal G}^2}{m^2T^2}+\frac{\pi}{2} \frac{8{\cal F}{\cal E}
    +{\cal G}^2}{mT^3} \right] \E^{-m/T}. \nonumber
\end{eqnarray}
These terms can be inserted into Eq. \re{2.14} together with the
classical, the one-loop, and the purely thermal part of the effective
Lagrangian, ${\cal L}_{\text{cl}}, {\cal L}^1$, and ${\cal L}^{1T}$.
Introducing the phase velocity $v=\frac{k^0}{|{\mathbf{k}}|}$ again
and omitting the terms of higher order in $T/m$ and $eB/m^2$, we
arrive at the following polarization sum rule describing the
modification of light propagation in a QED vacuum at low temperature
and weak magnetic background field:
\begin{eqnarray}
v^2&=&1-\frac{\alpha}{3}\sqrt{\frac{2}{\pi}} \sqrt{\frac{m}{T}}
\E^{-\frac{m}{T}} -\frac{22}{45} \frac{\alpha^2}{m^4} \, B^2
\sin^2\Theta \nonumber\\
&&-\frac{22}{45} \frac{\alpha^2}{m^4} \Biggl\{ \left[ \frac{\pi}{44}
  \frac{m^3}{T^3} -\frac{21}{22} \frac{m^2}{T^2} \right]
\E^{-\frac{m}{T}} \, B^2 \sin^2\Theta \nonumber\\
&&\qquad\quad+\left[\frac{2\pi}{11} \frac{m^3}{T^3} +\frac{4}{11}
    \frac{m^2}{T^2} \right]\E^{-\frac{m}{T}}\, B^2
  \Biggr\}. \nonumber\\
&&\label{2.14c}
\end{eqnarray}
In the first line of this equation, we encounter the familiar
classical, purely thermal, and purely magnetic contribution. The
second and third line describe the correction terms due to the
simultaneous presence of a heat bath and a magnetic background field,
indicating that the combination of both perturbations renders more
than a simple sum of the single effects. 

\section{Discussion}

\subsection{Low Temperature}

At temperatures well below the fermion mass, the one-loop modification
of the velocity of light as described by Eq. \re{2.9} vanishes
exponentially. It is the mass of the fermions in the loop that is
responsible for the attenuation of thermal effects, since it
counteracts thermal excitations of higher virtual modes. As a
consequence, the usual hierarchy of loop calculations is inverted: the
dominant thermal contribution to the low-temperature velocity shift
stems from the two-loop graph in which the internal photon line is
thermalized. Since the photon is massless, thermal excitations produce
a low-temperature velocity shift of $\delta v=-\frac{44\pi^2}{2025}
\alpha^2 \frac{T^4}{m^4}$ \cite{tarr83}-\cite{ditt98}. Incidentally,
the three-loop diagram involving two radiative thermalized photons of
course contributes subdominantly $\sim \alpha^3 \frac{T^8}{m^8}$, as
can be deduced from the findings of \cite{kong98}.

However, we would like to point out that the dominance of the two-loop
contribution does not seem to hold over the complete range where $T<
m$, as can be discovered numerically. The increasing influence of the
one-loop term is due to the factor of $T^{-1/2}$ and, of course, the
proportionality to $\alpha$ (and not $\alpha^2$) in Eq. \re{2.9}. The
one-loop contribution becomes dominant for comparably low temperatures
of $T/m\geq 0.058$ (Fig. 1). Of course, this statement should be
handled with care because here we are comparing a one-loop result
with thermalized fermions to a two-loop results without thermalized
fermions. Nevertheless, an increasing two-loop contribution from
thermalized fermions will always be suppressed by a factor of $\alpha$,
which could only be compensated for by an unexpected conspiration of
numerical pre-factors. 

\begin{figure}
\begin{center}
\epsfig{figure=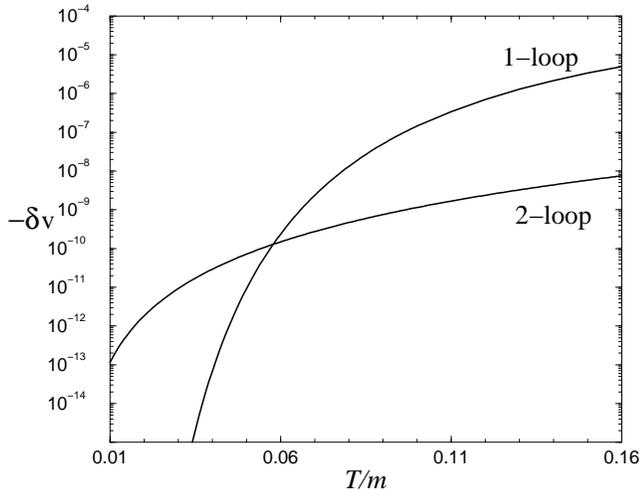,width=8cm}
\caption{Low-temperature velocity shift $\delta v$ ($\hat{=}
  \frac{1}{2}\delta v^2$) in units of the
  vacuum velocity $c=1$ versus the dimensionless temperature scale
  $T/m$; the one-loop contribution as given in Eq. \re{2.9} exceeds
  the well-known two-loop result $\sim$ $T^4/m^4$ for comparably low
  values of temperature.} 
\end{center}
\end{figure}

\subsection{High Temperature}

As discussed at the end of Sec. IV, the results of the formal
high-temperature expansion should be applied to values of temperature
of the order of the electron mass, $T\sim m$, where plasma effects
\cite{plasma} do not yet dominate the physical properties of the
modified vacuum. At even higher temperatures, $T\gg m$, the magnitude
of the plasma frequency $\omega_{\text{p}}$ serves as a cutoff for
the low-frequency waves which belong to the main ingredients of our
formalism. Then our calculations become meaningless, because the
low-frequency modes simply do not propagate.

Nevertheless, already in the temperature region where $T\sim m$, our
results indicate that the thermal excitations are no longer seriously
damped by the mass of the fermion in the loop. The one-loop
contribution obviously dominates the velocity shift; therefore, the
usual loop hierarchy is completely restored.

It is remarkable that a maximally possible velocity shift to this
order of calculation exists which is simply given by $\delta
v^2_{\text{max}}=-\frac{\alpha}{3\pi}$ (Fig. 2). Of course, this
maximal velocity shift is only reached asymptotically, and therefore,
strictly speaking, lies beyond the scope of the present formalism;
nevertheless, the actual velocity shift already comes close to the
maximum value in the intermediate-temperature domain where $T\sim m$
(Fig. 2). To complete the discussion, it should be mentioned that the
inclusion of the contributions from Eq. \re{2.7a} increase the
negative velocity shift proportional to $\frac{\alpha^2}{\pi^2} \ln
\frac{T}{m}$ for $T\geq m$. For reasonable values of temperature, this
contribution is in fact completely negligible.

\begin{figure}
\begin{center}
\epsfig{figure=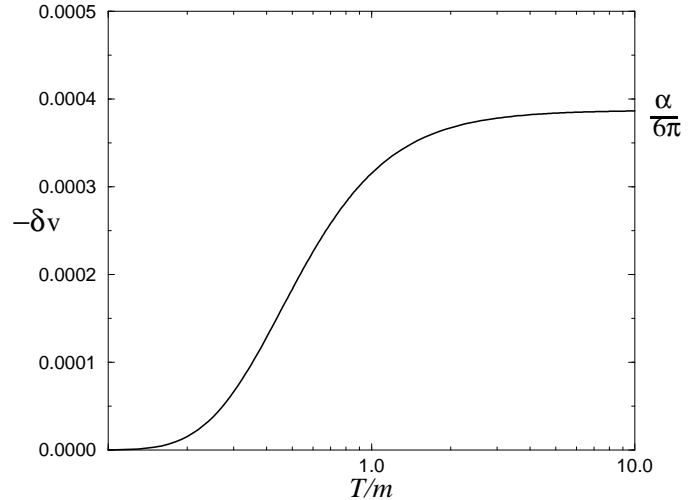,width=9cm}
\caption{Numerical evaluation of Eq. \re{2.6} in one-loop
  approximation: the thermally induced velocity shift
  $\protect{-{\delta} v}$ ($\hat{=}-\frac{1}{2} \delta v^2$) in units
  of the vacuum velocity $c=1$ is plotted versus the dimensionless
  temperature scale \mbox{$T/m$}; the left part of the curve is
  asymptotically described by Eq. \re{2.9}, the right part by Eq.
  \re{2.11}. The maximum velocity shift $-\delta
  v=\frac{\alpha}{6\pi}$ (equivalently: $-\delta
  v^2=\frac{\alpha}{3\pi}$) is already approached at comparably low
  temperatures.}
\end{center}
\end{figure}

\subsection{Low Temperature and Weak Magnetic Field}
In order to justify the low-temperature expansion leading to
Eq. \re{2.14b}, we restrict the following discussion of Eq. \re{2.14c}
to maximal values of temperature of
$T_{\text{max}}/m\simeq1/3$. 

Let us stress once more that the present results derive from a light
cone condition which represents a sum-rule for different polarization
modes. In fact, the QED vacuum under the influence of a oriented
background magnetic field is reminiscent of an anisotropic medium
with different refractive indices, i.e, velocity shifts, for each
polarization mode. In this sense, the velocity shifts found in
Eq. \re{2.14c} represent the arithmetic averages of those for the
single polarization modes. A detailed discussion of magnetically
induced birefringence at finite-temperature for which the eigenvalue
problem of the tensor $M^{\mu\nu}_{\alpha\beta}$ (Eq. \re{2.2a}) has
to be solved is out of the scope of the present work.

Concentrating on those terms in
Eq. \re{2.14c} which depend on the magnetic field, we encounter the
well-known zero-temperature velocity shift (last term of the first
line) with its typical dependence on the angle
$\Theta=<\!\!\!)({\mathbf{\hat{k},\hat{B}}})$; furthermore, there is
also one thermal correction with this $\Theta$-dependence (second
line), and another which is independent of the direction of the
magnetic field (third line). Although both thermal corrections
vanish exponentially in the zero-temperature limit, the factors of
$T$ in the numerator provide for a strong increase for $T/m>0.1$. 

Numerical analysis shows that the second line of Eq. \re{2.14c}
contributes with an opposite sign compared to the zero-temperature
part. Since both exhibit the same dependence on the magnetic field and
the angle $\Theta$, we may add them and find that, on the one hand,
thermal effects diminish the zero-temperature velocity shift in a
magnetic background by an amount of, e.g., $\simeq$20\% at
$T/m\simeq$0.25. On the other hand, the third line of Eq. \re{2.14c}
contributes with the same sign as the zero-temperature velocity shift,
and becomes comparable to the latter (for $\Theta=\pi/2$), e.g.,
$\simeq80\%$ of the zero-temperature velocity shift at
$T/m\simeq0.25$. Therefore, at these values of 
temperature, the velocity shift loses its strong dependence on the
direction of the magnetic field and the propagation direction and is
partly dependent on the energy density of the magnetic field.

\subsection{Effective Action Charge}
In Eq. \re{2.6b}, we were able to formulate the thermally induced
deformation of the light cone in terms of the vacuum expectation value
of the energy-momentum tensor times a factor $Q_T$ called effective
action charge \cite{ditt98}. The charge concept has been introduced as
a useful picture which provides for an intuitive understanding of this
proportionality factor. Since any perturbed QED vacuum can be expected
to control the magnitude of the velocity shifts even for vacuum
modifications of high energy density, this factor $Q_T$ has to
decrease sufficiently fast for increasing energy scale parameters (in
this case: temperature). Therefore, the factor $Q_T$ should exhibit a
localized distribution in this parameter space.

For calculating the effective action charge $Q_T$ according to Eq.
\re{2.6c} for a thermalized QED vacuum, we do not only consider the
one-loop contribution from the thermalized fermions ${\cal L}^{1T}$, but
also take the free photonic part ${\cal L}_\gamma=\frac{\pi^2}{45}
T^4$ into account. Although it does not exert an influence on the
velocity shift and drops out of Eq. \re{2.6b}, we have to include it
in the present considerations in order to work with the complete
vacuum expectation value of the energy-momentum tensor for the
thermalized QED vacuum.

To lowest order in $\alpha/\pi$, the formula for the effective action
charge Eq. \re{2.6c} reduces to:
\begin{equation}
Q_T\simeq\frac{2}{T} \frac{\partial_{\cal E}{\cal L}}{\partial_T{\cal
    L}} = \frac{2\partial_{\cal E}{\cal L}}{ T\partial_T {\cal
    L}_\gamma +T\partial_T {\cal L}^{1T}}, \label{3.15}
\end{equation}
where $\partial_{\cal E}{\cal L}^{1T}$ is given in Eq. \re{2.7}, and
the terms in the denominator can be written as:
\begin{eqnarray}
T\partial_T{\cal L}_\gamma (T)&=& \frac{4\pi^2}{45}\, T^4, \nonumber\\
T\partial_T{\cal L}^{1T}(T)&=& -\frac{2}{\pi^2} m^4 \frac{T}{m}
\sum_{n=1}^\infty \frac{(-1)^n}{n} \, K_3\bigl( {\scriptstyle
  \frac{m}{T}} n \bigr). \label{3.16}
\end{eqnarray}
Note that the appearance of $\partial_T{\cal L}_\gamma$ and
$\partial_T{\cal L}^{1T}$ in the denominator of Eq. \re{3.15}
corresponds to the appearance of a photonic and a fermionic part in
the energy-momentum tensor. One usually expects the fermionic part to
become important only for high temperature, $T\gg m$, where the 
fermions become ultra-relativistic.

\begin{figure}
\begin{center}
\epsfig{figure=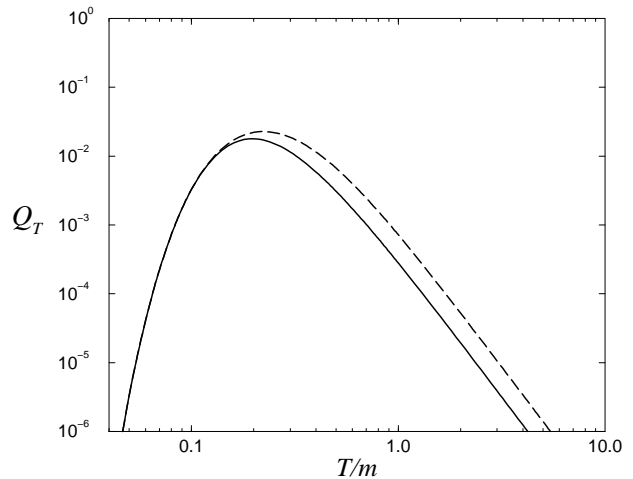,width=8cm}
\caption{One-loop contribution to the effective action charge $Q_T$ in
  units of $\frac{1}{m^4}$; for high temperature, $Q_T$ decreases
  proportional to $\frac{1}{T^4}$. The solid line corresponds to $Q_T$
  as given in Eq. \re{3.15}; for the dashed line, the fermionic
  contributions $\sim T\partial_T{\cal L}^{1T}$ have been omitted. In
  the low-temperature limit, the effective action charge to one loop
  vanishes due to the influence of a finite electron mass.} 
\label{figQT2}
\end{center}
\end{figure}
On the one hand, we indeed find the expected localized behavior, as
can be seen in Fig. 3: the effective action charge vanishes for high
as well as low temperatures; in between, it develops a maximum at $T/m
\simeq 0.22$. On the other hand, it is interesting to note that the
inclusion of the fermionic contributions $\sim T\partial_T{\cal
  L}^{1T}$ (solid line) in the denominator of Eq. \re{3.15} becomes
important for values of temperature close to the maximum of
$Q_T$. This again indicates that the thermalization of the fermions
becomes important already for comparably low values of
temperature.

Contrary to the cases discussed in \cite{ditt98}, the one-loop
effective action charge is not centered at the origin. This is because
the electron mass damps the fermionic thermal fluctuations
exponentially for small values of temperature.

\section{Concluding Remarks}

First, we would like to stress that the one-loop contributions to the
velocity shift as calculated in the present work do not fit into the
scheme of the ``unified formula'' proposed in \cite{lato95}; the
latter connects the velocity shift with a shift of the vacuum energy
density caused by external influences with a universal constant
coefficient as proportionality factor. In the language of Eq.
\re{2.6b}, this coefficient has to be identified with the effective
action charge $Q_T$ which in the present case is not constant at all
but almost carries the complete physical information of the problem.
This misfit indicates that the ``unified formula'' is an artefact of
an approximation scheme rather than a fundamental principle
\cite{ditt98}.

Furthermore, it should be pointed out that the maximum velocity shift,
$-\delta v^2_{\text{max}}=\frac{\alpha}{3\pi}$, cannot be viewed as an
experimentally significant limiting value, since light propagation
will successively be dominated by plasma effects for increasing
temperature at $T\sim m$. Nevertheless, the existence and amount of
such a maximum shift are at least interesting from a theoretical
viewpoint, since they characterize the classically forbidden
interaction between the modified vacuum and a photon which is exposed
to the effects of vacuum polarization. Supposing there were no
collective excitations constituting a plasmon for high temperature,
then QED would not allow for an arbitrarily strong influence of a heat
bath on the propagation of light for reasonable values of temperature.
Of course, for extremely high temperature, the logarithmic corrections
from $\partial_{\cal F}{\cal L}$ in Eq. \re{2.6} would also
significantly increase the negative velocity shift, slowing down the
speed of light; e.g., another shift of $\alpha/3\pi$ would be reached
at $T/m\simeq 6.4\cdot 10^{7}$.

The problem of calculating the intermediate-/high-temperature velocity
shift was also approached in \cite{ditt98} employing a method
reminiscent of the Born-Oppenheimer approximation: the coupling of the
external field to the fermion loop was considered as an adiabatic
interaction, while the photonic fluctuations were treated as the
``fast'' degrees of freedom.  This calculation can be rated as a
two-loop calculation, since the thermalization of the fermions has
been approximately taken into account by using a thermal one-loop
Lagrangian; photonic fluctuations were introduced in the same way as
in the low-temperature case, i.e., by taking thermal vacuum
expectation values of the field quantities.  Similarly to the present
work, a maximum velocity shift has been found which is about three
orders of magnitude smaller than the one-loop result given above.
Indeed, this fits into the usual hierarchy of perturbation theory.

To overcome the approximation scheme of \cite{ditt98}, which treats the
thermalization of fermions and photons on a different footing, the
present paper offers an appropriate tool by working with the complete
set of invariants of the thermalized QED vacuum. However, for a direct
verification of the results of \cite{ditt98} for high temperature, a
first-principles two-loop calculation of the effective action should be
carried out in the same sense as suggested in the present work. 

Beyond the clarification of these technical questions, a two-loop
calculation of the thermal effective action will be useful for
studying further phenomenological aspects of the physics of strong
fields, such as thermally induced pair production or photon
splitting. In view of the present results, one might speculate that,
for these effects, two-loop contributions will be dominant at low
temperature, since thermal one-loop corrections have proved to be
vanishing (for pair production \cite{cox84,elmf94,gies98}) or of minor
importance (for photon splitting \cite{elmf98}). 

\acknowledgements

I would like to thank Prof. W. Dittrich for insightful discussions and
for carefully reading the manuscript. I am especially grateful to Dr.
K. Scharnhorst for encouraging discussions and for his detailed
constructive criticism. Valuable comments by Prof. G. Barton are also
thankfully acknowledged.

\end{document}